\documentclass[prc,aps,preprint,superscriptaddress]{revtex4}
\usepackage{graphicx}
\begin{document}

\preprint{INT-PUB-10-008}
\preprint{LLNL-JRNL-423216}
\title{Sensitivity analysis of random two-body interactions }

\author{Calvin W. Johnson}
\affiliation{Department of Physics, San Diego State University,
5500 Campanile Drive, San Diego, CA 92182-1233}
\author{Plamen G. Krastev}
\affiliation{Department of Physics, San Diego State University,
5500 Campanile Drive, San Diego, CA 92182-1233}
\affiliation{Lawrence Livermore National  Laboratory, P.O. Box 808, L-414, 
Livermore, CA 94551}

\begin{abstract}
The input to the configuration-interaction shell model includes many dozens 
or hundreds of 
independent two-body matrix elements. Previous studies have 
shown that when fitting to experimental low-lying spectra, the greatest sensitivity is to only a few 
linear combinations of matrix elements. Here we consider interactions drawn from the 
two-body random ensemble, or TBRE,  and find that the low-lying spectra are also most sensitive to only a few linear combinations of two-body matrix elements, in a fashion nearly indistinguishable from an interaction empirically fit to data. 
We find in particular the spectra for both the random and empirical interactions are 
sensitive to similar matrix elements, which we analyze using monopole 
and contact interactions. 
\end{abstract}
\maketitle

\section{Introduction and Motivation}

The configuration-interaction shell model is a useful framework for a detailed understanding of low-energy 
nuclear structure \cite{BG77,br88,ca05}. The many-body basis is a large dimension ($10^{3-10}$) set of Slater determinants, 
which are antisymmeterized products of single-particle states. One must truncate the
 single-particle states, corresponding to one or a few  shells 
(typically using the harmonic oscillator as an approximation to the mean-field); 
the many-body basis may be further truncated. 
For phenomenological calculations one writes the Hamiltonian in terms of single-particle 
energies and two-body matrix elements, 
while for \textit{ab initio} calculations one may extend this to 
three-body interactions \cite{NO03}.

The two-body matrix elements  are the matrix elements of the residual interaction in the 
lab frame, 
\begin{equation}
V_{JT}(ij,kl) = \langle ij; JT| \hat{V} |kl; JT \rangle
\end{equation}
where $|ij; JT \rangle$ is the normalized, antisymmeterized product of particles in orbits labeled by 
$i$ and $j$ and coupled to good angular momentum $J$ and isospin $T$. 
 If one starts from a translationally invariant interaction between particles, 
one can either compute the integral in the lab frame or start in the relative frame 
and then transform to the lab frame; in either case 
there are correlations between the matrix elements, 
although they are not obvious to the casual observer.

Often for semi-phenomenological calculations, 
one starts from a ``realistic'' interaction, and then adjusts the 
two-body matrix elements until the rms error on a set of 
experimentally known energy levels is minimized \cite{BG77}.   
In the $1s$-$0d$ shell, such a  semi-phenomenological interaction 
has been recently derived \cite{br06}, 
improving on an earlier interaction\cite{Wildenthal}.

It has been found that the fits are empirically 
dominated by a few linear combinations of matrix elements. 
The physical meaning of those dominant combinations is not 
immediately obvious. One might naively guess the linear correlations are due to 
an underlying translationally invariant interaction 
(although a density dependence would destroy this).
Somewhat more phenomenologically, it has been argued by appealing to mean-field 
properties that one 
can improve fits primarily through adjusting the monopole-monopole part of the interaction,
that is, interaction terms that look like  $n_a (n_b - \delta_{ab})$, where $n_a$ is 
the number of particles in the $a$th orbit.  This protocol for shifting 
monopole strengths has been successfully applied to several semi-empirical 
interactions \cite{po81,ma97,ut99,ho04,su06} 

A related study \cite{BJ09}, investigating the origin of many-body forces from 
truncation of the model space, also found an empirical fit dominated by a 
few linear combinations of matrix elements. While much of the fit was dominated 
by the monopole interactions, even better agreement was brought about  using 
a contact interaction motivated by its usage in mean-field calculations \cite{sk56,be03}
and effective field theory\cite{vk99}.

In investigating the character and origin of the dominant matrix elements, it is 
useful to ask if there is anything special about the nuclear interaction. One way to 
ask this question is to compare with interactions drawn from the two-body random ensemble (TBRE), 
which despite their arbitrary nature are known to echo some features of real nuclear spectra 
\cite{JBD98,ZV04,ZAY04,PW07}. 
In this paper we conduct a sensitivity analysis of the low-lying spectra of 
random interactions and compare against a standard empirical interaction, USDB. 
We find that for more measures all the interactions are nearly indistinguishable, 
at least on a statistical level.

\section{Methodology and results}

Our methodology follows previous work \cite{BG77,br06,BJ09}; we 
work in the $1s$-$0d$ valence space with an inert $^{16}$O core.
Given an input set of two-body matrix elements (we leave aside single-particle 
energies and any $A$-dependent scaling), which we write as a vector $\vec{v}$, 
we can calculate the eigenvalues $E_\alpha(\vec{v})$ of the many-body Hamiltonian.
For this work the label $\alpha$ ranged over 
all nuclides with $ 0 \leq Z_\mathrm{valence} \leq N_\mathrm{valence} \leq 10$ and 
took the ground state binding energy and the first five excitation energies. 

If one has a target spectrum $E_\alpha^0$, say from experiment, then the goal 
of the fit is to minimize
\begin{equation}
\sum_\alpha \left( E_\alpha(\vec{v}) - E_\alpha^0\right )^2 \label{chi2}
\end{equation}
(for simplicity we leave off the experimental uncertainty in each state). 
Expanding to first order
\begin{equation}
 E_\alpha(\vec{v}+ \delta\vec{v}) \approx  E_\alpha(\vec{v}) + \sum_i \delta v_i 
\frac{\partial E_\alpha}{\partial v_i}
\end{equation}
then minimizing (\ref{chi2}) yields
\begin{equation}
\sum_\alpha \left( E_\alpha(\vec{v}) - E_\alpha^0\right )
\frac{\partial E_\alpha}{\partial v_i} = 
\sum_j \sum_\alpha  \frac{\partial E_\alpha}{\partial v_i} 
\frac{\partial E_\alpha}{\partial v_j} \delta v_j.
\end{equation}
This equation is in the form $\vec{b} = \mathbf{A} \vec{x}$ 
where 
\begin{equation}
A_{ij} = \sum_\alpha  \frac{\partial E_\alpha}{\partial v_i} 
\frac{\partial E_\alpha}{\partial v_j} . \label{defA}
\end{equation}
The derivatives come via the Hellmann-Feynman theorem \cite{F39}
\begin{equation}
  \frac{\partial E_\alpha}{\partial v_i} = 
\left \langle \Psi_\alpha \left | \hat{H}_i \right| \Psi_\alpha \right \rangle
\end{equation}
where $\hat{H}_i$ is the Hamitonian operator whose strength is $v_i$. 

We then find the eigenvalues of $\mathbf{A}$ (which is equivalent to finding 
the squares of the eigenvalues in the singular-value decomposition of 
$\partial E_\alpha/\partial v_i$).   We do this for both the USDB interaction
and for an ensemble of 100 sets of random two-body interactions, also called 
the two-body random ensemble (TBRE). 
The results are shown for the TBRE in Fig.~1, where we have separated out the 
sensitivity just for the binding energies (ground state energies) and the 
excitation energies. Although not shown, the equivalent SVD eigenvalues for 
USDB are completely within the TBRE results. 

\begin{figure}  
\includegraphics [width = 7.5cm]{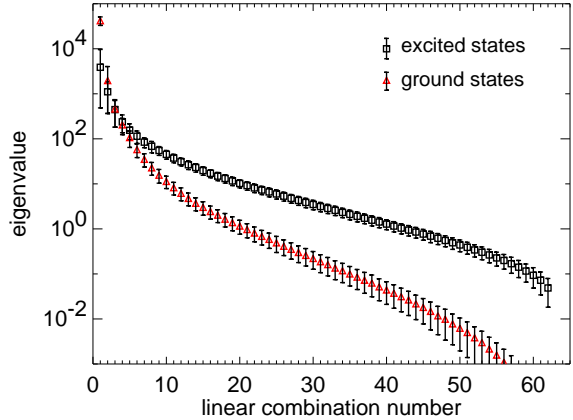}
\caption{\label{svd_spectra} (Color online)The spectra of eigenvalues 
from a singular-value decomposition of the sensitivity 
matrix $\mathbf{A}$ (Eq.~\ref{defA}) for the two-body random ensemble. 
The lower curve is for 
ground state energies, while the upper curve is for excitation energies.
}
\end{figure}

The lower curve is for ground states only, while the upper curve is for excitations 
energies relative to the ground state.  Clearly, and 
perhaps unsurprisingly, the ground state energies are predominantly sensitive to 
just a few linear combinations of matrix elements--significantly fewer than 
excitations energies. 

Fig.~1 characterizes \textit{eigenvalues} of $\mathbf{A}$. The next step is 
to characterize the \textit{eigenvectors} associated with the dominant 
eigenvalues, specifically by comparing with monopole and contact 
interactions.  To do so we first discuss a method for quantifying the 
overlap between two vector subspaces \cite{MC, SK08}. Consider two vectors subspaces,
$S_1$ and $S_2$. Let $\mathbf{V}_1$ be a matrix whose column vectors are the 
(orthonormal) basis vectors of $S_1$, and similarly with $\mathbf{V}_2$. From these 
one constructs the overlap matrix $\mathbf{\Omega} = \mathbf{V}_1^\dagger \mathbf{V}_2$. 
Note that if the subspaces are not of equal dimension then $\mathbf{\Omega}$ is 
not a square matrix. In any case we do a singular value decomposition of $\mathbf{\Omega}$;
the SVD eigenvalue spectrum then is a measure of the overlap of the two spaces.  
If the two spaces perfectly overlap then all eigenvalues are 1, if just $N$ of the 
dimensions perfectly overlap than $N$ eigenvalues will be 1 and the rest zero.  
Note that this method is invariant under arbitrary choice of orthonormal bases. 

We begin with the monopole-monopole interaction of the form $n_a (n_b -\delta_{ab})$, 
which has six unique terms, and thus six vectors or linear combinations 
of matrix elements, in the $sd$-shell. These we combine with the $k$ most dominant 
linear combinations that arise from the previous analysis; somewhat arbitrarily we 
chose $k=6$ (our results do not change qualitatively for other small values of $k$). 
The results, the SVD eigenvalues of $\Omega$, are shown in Fig.~2.  It is important to 
note we compute the spectrum separately for each randomly generated interaction 
and then compute the distribution.

\begin{figure}  
\includegraphics [width = 7.5cm]{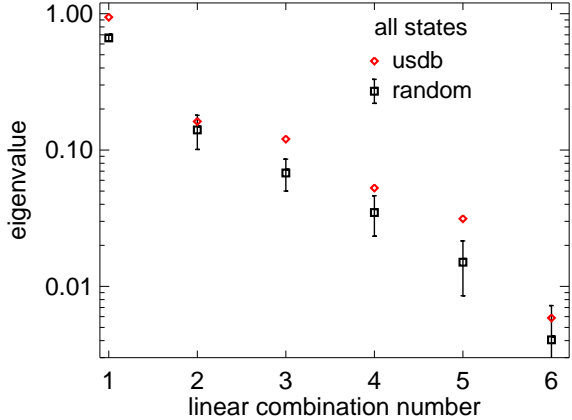}
\caption{\label{overlap_mono} (Color online) The SVD spectrum that measures 
the overlap of the subspace defined by the six largest eigenvectors 
from Fig.~1, with the subspace defined by the monopole-monopole 
interaction. (Black) squares are for the TBRE, while (red) diamonds are 
for USDB. 
}
\end{figure}

The results for ground states and for excited states are similar, so we combine 
all states into a single calculation. The eigenvalues for USDB are roughly $50\%$ higher 
than for the TBRE, but otherwise qualitatively very similar.

We also compared for contact interactions; we took only two terms, the 
$S=0, T=1$ channel and $S=1, T=0$ channel (there being only the $s$-wave channel 
in relative coordinates). These results we summarize in Table 1.  
\begin{table}[h!]
\caption{Leading eigenvalues from SVD of subspace overlaps of two-term contact
interaction
}
\begin{tabular}{|c|ccc|}
\colrule
Interaction & ground states & excited states & all states \\
\colrule 
USDB & 0.60 &  0.58  &  0.62 \\
TBRE & $0.55 \pm 0.04$ &  $ 0.41 \pm 0.06 $ & $0.55 \pm 0.04$ \\
\colrule
\end{tabular}
\end{table}

For comparison, the leading eigenvalue for the overlaps of USDB versus the six-term 
monopole is 0.94, while that of the TBRE versus monopole is $0.66 \pm 0.03$. 
There is somewhat 
more sensitivity to the monopole interaction than the contact interaction; however, 
the reader should keep in mind that is not the whole story. Recall that when fitting 
an interaction, the linearized equations are cast in the form $\mathbf{A}\vec{x} = 
\vec{B}$. Our analysis in this paper is entirely with the eigenvalues of $\mathbf{A}$, 
but in any fit one must also look to $\vec{b}$ (which in practice is the deviation of the 
theoretical spectra from experiment). For example, in \cite{BJ09} it was found that 
using a contact interaction brought better agreement than a monopole interaction. One 
can understand this in terms of conjugate gradient methods \cite{NR}: the direction 
of local steepest gradient may not in fact point towards the global minimum. 

\begin{figure}  
\includegraphics [width = 7.5cm]{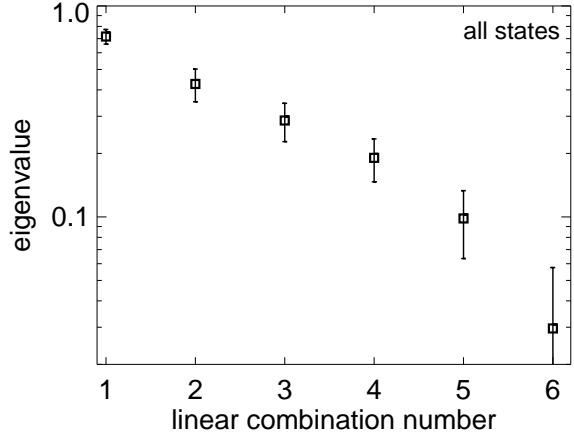}
\caption{\label{usdb_random} SVD spectrum from the overlap of the dominant eigenvectors 
from USDB and the TBRE.
}
\end{figure}

By our measures so far, both the TBRE and the empirically-fit USDB look qualitatively
similar. Therefore we take a final analysis by comparing the dominant 
linear combinations of the USDB with those from the TBRE.  This is show in 
Fig.~3, using the same analysis as for Fig.~2.  For comparison with the previous 
results, the leading eigenvalue is  $0.72 \pm 0.06$.

\section{Conclusion}

We have analyzed the sensitivity of the low-lying spectra of the random two-body ensemble of interactions to variations of the Hamiltonian matrix elements; by using singular value decomposition, we find 
the dominant linear combinations, which would be important in any fit to experimental data. 
 We found the SVD eigenvalues follow a pattern remarkably similar to that shown by 
semi-realistic/semi-phenomenological interactions such 
as USDB. We also analyzed the most dominant linear combinations of matrix elements by computing the 
overlap with monopole and contact interactions. Overall, both the TBRE and the empirical USDB had 
qualitatively similar results.

The U.S.~Department of Energy supported this investigation through
contracts DE-FG02-96ER40985 and DE-FC02-09ER41587, and through subcontract B576152 by Lawrence Livermore National 
Laboratory under contract DE-AC52-07NA27344.

\end{document}